\documentclass[%
reprint,
superscriptaddress,
amsmath,amssymb,
aps,
prb,
floatfix,
]{revtex4-2}

\usepackage{graphicx}
\usepackage{dcolumn}
\usepackage{bm}
\usepackage{xcolor}
\usepackage{float}
\usepackage{amsmath}

\usepackage[normalem]{ulem} 
\usepackage{color}
\definecolor{red}{rgb}{1,0,0}
\definecolor{blue}{rgb}{0,0,1}

\begin{document}

\title{Voltage-amplified heat rectification in SIS junctions}
\author{Ilia Khomchenko}\email{ilya.khomchenko@skoltech.ru}
\affiliation{Skolkovo Institute of Science and Technology, 3 Nobel Street, Skolkovo, Moscow Region 121205, Russia}
\affiliation{Center for Nonlinear and Complex Systems, Dipartimento di Scienza e Alta Tecnologia, Università degli Studi dell’Insubria,
via Valleggio 11, 22100 Como, Italy}
\affiliation{Istituto Nazionale di Fisica Nucleare, Sezione di Milano, via Celoria 16, 20133 Milano, Italy}
\author{Henni Ouerdane}\email{h.ouerdane@skoltech.ru}
\affiliation{Skolkovo Institute of Science and Technology, 3 Nobel Street, Skolkovo, Moscow Region 121205, Russia}
\author{Giuliano Benenti}\email{giuliano.benenti@uninsubria.it}
\affiliation{Center for Nonlinear and Complex Systems, Dipartimento di Scienza e Alta Tecnologia, Università degli Studi dell’Insubria,
via Valleggio 11, 22100 Como, Italy}
\affiliation{Istituto Nazionale di Fisica Nucleare, Sezione di Milano, via Celoria 16, 20133 Milano, Italy}
\affiliation{NEST, Istituto Nanoscienze-CNR, I-56126 Pisa, Italy}

\date{\today}

\begin{abstract}
The control of thermal fluxes -- magnitude and direction, in mesoscale and nanoscale electronic circuits can be achieved by means of heat rectification using thermal diodes in two-terminal systems. The rectification coefficient $\mathcal{R}$, given by the ratio of forward and backward heat fluxes, varies with the design of the diode and the working conditions under which the system operates. A value of $\mathcal{R}\ll 1$ or $\mathcal{R}\gg 1$ is a signature of high heat rectification performance but current solutions allowing such ranges, necessitate rather complex designs. Here, we propose a simple solution: the use of a superconductor-insulator-superconductor (SIS) junction under an applied fast oscillating (THz range) voltage as the control of the heat flow direction and magnitude can be done by tuning the initial value of the superconducting phase. Our theoretical model based on the Green functions formalism and coherent transport theory, shows a possible sharp rise of the heat rectification coefficient with values up to $\mathcal{R} \approx 500$ beyond the adiabatic regime. The influence of quantum coherent effects on heat rectification in the SIS junction is highlighted.
\end{abstract}

\maketitle

\section{Introduction}
\label{sec:intrp}

Quantum technology is the pathway to breakthroughs in sectors such as metrology, communication, big data, sensing, and computing to name a few owing to the increasing ability to harness the properties of quantum systems \cite{dowling2003quantum, riedel2017european, acin2018quantum, fedorov2019quantum,qcbook}. In computing and data science, a critical step is the design and implementation of high-performance quantum processors \cite{preskill2012quantum, harrow2017quantum}, which would in turn stimulate the wide-scale development of quantum computers \cite{ladd2010quantum, dyakonov2019will,castelvecchi2017quantum, fedorov2018quantum, arute2019quantum, wu2021strong}. One of the many challenges on the path to this goal, notably for technologies based on solid-state qubits, is the control of heat transport on the mesoscopic and nanoscopic scales \cite{giazotto2006opportunities,LiLi2012,pekola2015towards,fornieri2017towards,pekolakarimi2021}. A possible approach to that problem is to use thermal diodes as circuit components that can control the direction and magnitude of heat currents \cite{BenentiPeyrard2016,pekolakarimi2021}. Thermal diodes act as heat rectifiers since their thermal conductance is asymmetric along the direction of the temperature gradient and they may also find applications for, e.g., energy harvesting and radiation detection \cite{giazotto2006opportunities}. The efficiency of a heat rectifier is given by the rectification coefficient, a quantitative indicator of the amount of heat that can be transferred or blocked in a forward or a backward direction. 

Although heat rectification was experimentally observed 85 years ago in the pioneering work of Starr \cite{starr1936copper} and subsequently studied in many works, mostly from the years 2000s \cite{terraneo2002controlling,chang2006solid,scheibner2008quantum,fornieri2014normal,dettori2016thermal,balachandran2018perfect,balachandran2019,marchegiani2021highly}, the first experimental realizations using superconducting materials and their detailed studies are fairly recent~\cite{giazotto2013,martinez2014coherent,fornieri2015electronic,martinez2015rectification,senior2020heat,giazotto2020very}. Typically the systems are junctions made of a combination of normal metal, insulating, and superconducting materials. Heat rectification in these systems can be very efficient owing to the presence of superconducting tunneling junctions that foster the enhancement of the thermal asymmetry. 
Indeed, a superconductor acts like a low-pass filter for charges whose energy is smaller than the temperature-dependent gap $\Delta(T)$~\cite{martinez2015rectification}. The asymmetry in a SIS junction is obtained if the two (left and right) superconductors have different energy gaps, say $\Delta_L(T)>\Delta_R(T)$. The heat flux in the forward configuration ($T_L>T_R$) is maximized when there is a matching of singularities in the superconducting density of states, that is, for $\Delta_L(T_L)=\Delta_R(T_R)$. Such matching is not possible in the reverse configuration ($T_L<T_R$), and such asymmetry leads to heat rectification.

In the works mentioned above, heat rectification, while driven by the externally-imposed temperature bias, essentially depends on the intrinsic and combined properties of the materials that are used to make the thermal diode. Here, we theoretically investigate a possible sizable increase of the rectification of heat carried by electronic currents by means of an external voltage bias, which affects the superconducting phase \cite{josephson1964coupled}. In a more general perspective, the time evolution of heat fluxes under alternating voltages is an essential problem to treat \cite{larkin1967tunnel} \footnote{For phase-dependent thermal transport in Josephson junctions in the absence of a bias voltage, 
see~\cite{maki1965entropy, maki1966entropy,zhao2003phase, zhao2004heat}}. In this paper, considering a simple
SIS junction, we show that an applied oscillating voltage can boost the heat rectification coefficient to large values, up to $\mathcal{R}\approx 500$, which is comparable in orders of magnitude to the level of performance that more complex hybrid thermal diodes can boast \cite{martinez2015rectification}. 

\section{Model of heat currents in a Josephson junction}

\label{sec:model}

\begin{figure}[h]
  \centering
  \includegraphics[width=.45\textwidth]{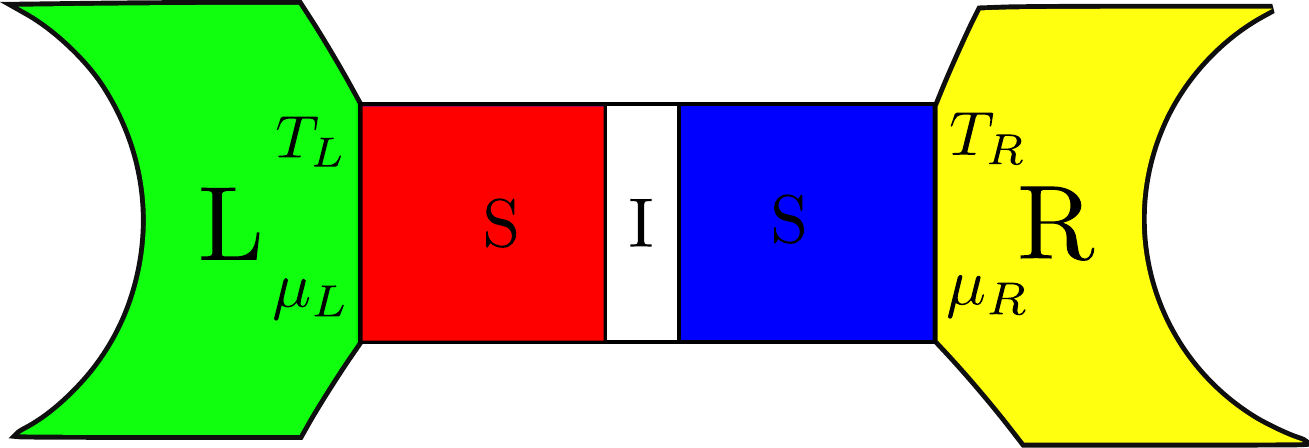}
  \caption{Schematic representation of a SIS junction coupled to the reservoirs.}
  \label{figone}
\end{figure}

We consider a SIS junction weakly coupled to reservoirs as illustrated in Fig.~\ref{figone}. Both reservoirs are characterized by their temperatures $T_{\rm L}, T_{\rm R}$ and electrochemical potentials $\mu_{\rm L}, \mu_{\rm R}$, respectively. The condition of the weak coupling is fulfilled when the Dynes parameter, $\Gamma$, which is related to the finite lifetime of the quasiparticle states, is small compared to the superconducting energy gap: $\Gamma \ll \Delta$. Note that though we consider the weak coupling regime, we assume that the coupling of the superconductors to the insulating barrier remains sufficiently strong to neglect phase changes across the barrier~\cite{josephson1964coupled}. The condition of the sufficiently strong coupling with the barrier implies that the coupling energy is comparable to $k_{\rm B}T_{\rm c}$, with the $T_{\rm c}$ being the superconducting transition temperature and $k_{\rm B}$ the Boltzmann constant. Denoting $e$, the elementary charge, the voltage $V=1/e(\mu_{\rm L}-\mu_{\rm R})$ across the junction is related to the superconducting phase $\varphi(t)$ via the Josephson equation:
\begin{equation}
\frac{{\rm d} \varphi}{{\rm d} t} = \frac{2eV}{\hbar},
\label{eq:1}
\end{equation}
where $\hbar$ is the reduced Planck constant.

The heat current through the tunnel junction can flow from left to right, which we call the forward (fw) direction, and from right to left, or backward (bw) direction. In this work, we assume $T_{\rm L}> T_{\rm R}$ for the forward flux, and $T_{\rm L} < T_{\rm R}$ for the backward heat flux. The contributions to the heat current $\dot{Q}$ are due to the following: the active and reactive parts of the quasiparticle (or single-electron) charge current, $\dot{Q}_{\mathrm{qp}}$ and $\dot{Q}_{\mathrm{qpr}}$ respectively; the Cooper pairs tunneling through the junction, $\dot{Q}_{\rm j}$; and the interference because of Cooper pairs breaking and formation on the electrodes of the junction, $\dot{Q}_{\rm int}$. These contributions have been studied in detail \cite{maki1965entropy, maki1966entropy, guttman1997phase}, and a convenient and simple formalism for heat fluxes in general cases can be found in~\cite{barone1982physics}, the key idea being to introduce the spectral decomposition for the phase factor:
\begin{equation}
	e^{-j\varphi/2} = \int_{-\infty}^{\infty} \frac{{\rm d} \omega}{2 \pi} W(\omega)e^{-j \omega t},
\label{eq:2} 
\end{equation}
where $W(\omega)$ denote the Fourier coefficients and $j^2=-1$. For simplicity, we assume that the Fourier coefficients are real. Adopting a similar approach as for the derivations for charge currents~\cite{ambegaokar1963tunneling, werthamer1966nonlinear, larkin1967tunnel, harris1975josephson, barone1982physics}, we obtain the following expressions for the heat fluxes in the directions $i \equiv \mathrm{fw},\mathrm{bw}$:
\begin{eqnarray}
\label{eq:3} 
&&\dot{Q}_{i}(t) = \int_{-\infty}^{\infty} {\rm d} \omega \int_{-\infty}^{\infty} {\rm d} \omega' W(\omega) W(\omega')\\
\nonumber
&& \times\left[\dot{Q}_{{\rm qp};i}(\hbar \omega')\cos(\omega-\omega') t - \dot{Q}_{{\rm qpr};i}(\hbar \omega')\sin(\omega-\omega') t\right.\\ 
\nonumber 
&& + \ \dot{Q}_{{\rm int};i}(\hbar \omega')\cos( \varphi_{0}+ (\omega+\omega') t)\\
\nonumber 
&& + \left. \dot{Q}_{{\rm j};i}(\hbar \omega')\sin(\varphi_{0}+ (\omega+\omega') t)\right], 
\end{eqnarray}
where $\varphi_{0} = \varphi_{L} - \varphi_{R}$ is the initial phase difference between the two superconductors \cite{harris1975josephson, barone1982physics}. The tunneling contribution $\dot{Q}_{\rm j}$ and reactive contribution $\dot{Q}_{\mathrm{qpr}}$ are symmetric with respect to temperature reversal; conversely, the quasiparticle $\dot{Q}_{\mathrm{qp}}$ and the interference contributions $\dot{Q}_{\mathrm{int}}$ are antisymmetric; their expressions are given in Appendix~\ref{AppA}.
These formulas are valid for any frequency and temperature~\cite{larkin1967tunnel,barone1982physics}. 

The heat rectification coefficient $\mathcal{R}$ is given by
\begin{equation}
    \mathcal{R} = \frac{\dot{Q}_{\mathrm{fw}}}{\dot{Q}_{\mathrm{bw}}}
\label{eq:4}     
\end{equation}
and, in this work, we analyze its behavior under two regimes of the externally applied voltage: adiabatic and nonadiabatic.
In the adiabatic regime, the potential difference $V(t)\equiv V_{\rm s}(V_{{\rm s}0},\Omega)$, with $V_{\rm s0}$ being the amplitude such that $eV_{{\rm s}0} \ll \Delta$, and $\Omega$ the frequency such that $\hbar \Omega \ll \Delta$~\cite{larkin1967tunnel, de1983time}; and the expressions for the heat currents are significantly simplified :
\begin{equation}
\dot{Q}_{i} = \pm \dot{Q}_{\mathrm{qp};i} + \dot{Q}_{{\rm j};i}\sin\varphi \pm \dot{Q}_{{\rm int};i} \cos\varphi, 
\label{eq:5}
\end{equation}
where $\pm$ is for the contributions that are antisymmetric with respect to the temperature gradient~\cite{martinez2014coherent}. 

Under an arbitrary voltage drop $V(t)$, the theoretical framework developed by Larkin and Ovchinnikov \cite{larkin1967tunnel} and independently by Werthamer \cite{werthamer1966nonlinear}, yields quite cumbersome formulas, which makes the analysis of heat rectification rather involved. However, valuable insights beyond the adiabatic regime can be obtained if one considers a voltage drop of the form: $V(t) = V_{\rm s}(t)+V_{\rm r}(t)$, with $V_{\rm s}(t)$ being the slowly-varying part with frequency $\Omega$, and $V_{\rm r}(t)$ the rapidly-varying part with frequency $\omega_0$ violating the adiabaticity condition. In this case we obtain~\footnote{Following the derivation of Refs.~\cite{larkin1967tunnel, de1983time}, we obtain the forward and the backward heat fluxes using the same sign convention as in \cite{werthamer1966nonlinear, harris1974cosine, harris1975josephson, barone1982physics, de1983time} rather than that in \cite{larkin1967tunnel}.}

\begin{widetext}
\begin{eqnarray}
\label{eq:6} 
\dot{Q}_{i}(t) &=&  \int_{-\infty}^{\infty}\! \frac{d \omega}{2 \pi \hbar \omega} \bigg\{ \mathrm{Re}  \left(eV_{\rm r}(\omega) \exp\left[ -j \left(\omega t + \int^{t} \frac{2e V_{\rm s}(t)}{\hbar} {\rm d}t \right) \right]\right) 
\left[\dot{Q}_{\mathrm{int};i}(V_{{\rm s}0} + \hbar \omega/e)-\dot{Q}_{{\rm int};i}(V_{{\rm s}0})\right]
\\ \nonumber 
&+&  \mathrm{Im}  \left(eV_{\rm r}(\omega) \exp \left[ -j \left( \omega t + \int^{t} \frac{2eV_{\rm s}(t)}{\hbar} {\rm d}t \right) \right] \right) 
\left[\dot{Q}_{{\rm j};i}(V_{{\rm s}0} + \hbar \omega/e)-\dot{Q}_{{\rm j};i}(V_{{\rm s}0})\right]
\\ \nonumber 
&+&  \mathrm{Re} \left( eV_{\rm r}(\omega) \exp (-j\omega t)\right) \left[\dot{Q}_{\mathrm{qp};i}(V_{{\rm s}0} +  \hbar \omega/e)-\dot{Q}_{\mathrm{qp};i}(V_{{\rm s}0})\right] 
\\ \nonumber 
&+&  \mathrm{Im} \left( eV_r(\omega) \exp (-j\omega t) \right) \left[\dot{Q}_{\mathrm{qpr};i}(V_{{\rm s}0} + \hbar \omega/e)-\dot{Q}_{\mathrm{qpr};i}(V_{{\rm s}0})\right] \bigg\},
\end{eqnarray}
\end{widetext}
which is valid on the condition that $eV_{\mathrm{r}} \ll \hbar \omega_{0}$~\cite{larkin1967tunnel} \footnote{Note that throughout this work, we ignore the spatial dependence of the superconducting phase, which can be taken into account following~\cite{werthamer1966nonlinear}}. Now, taking $V_{\rm r}(t) = V_{{\rm r}0}\mathrm{cos} (\omega_{0}t)$, its Fourier transform reads $V_{\rm r}(\omega) = V_{{\rm r}0} \pi [\delta(\omega - \omega_{0})+\delta(\omega + \omega_{0})]$, and Eq.~\eqref{eq:6} can be simplified to
\begin{equation}
\dot{Q}_{i}(t) = \dot{Q}_{i}(\omega_{0}t) + \dot{Q}_{i}(-\omega_{0}t),
\label{eq:7}   
\end{equation}
with 
\begin{eqnarray}
\label{eq:8}    
&&\dot{Q}_{i}(\omega_{0}t) =  \\
\nonumber 
&&{}\frac{eV_{{\rm r}0}}{\hbar \omega_{0}}\left[\dot{Q}_{\mathrm{int};i}(V_{{\rm s}0} + \hbar \omega_{0}/e)-\dot{Q}_{\mathrm{int};i}(V_{{\rm s}0})\right]\cos(\varphi_{\rm s}(t)-\omega_{0}t) 
\\ 
\nonumber 
&&{}+ \frac{eV_{{\rm r}0}}{\hbar \omega_{0}}\left[\dot{Q}_{{\rm j};i}(V_{{\rm s}0} + \hbar \omega_{0}/e)-\dot{Q}_{{\rm j};i}(V_{{\rm s}0})\right]\sin(\varphi_{\rm s}(t)-\omega_{0}t) 
\\ 
\nonumber
&&{}+ \frac{eV_{{\rm r}0}}{\hbar \omega_{0}}\left[\dot{Q}_{\mathrm{qp};i}(V_{{\rm s}0} + \hbar \omega_{0}/e)-\dot{Q}_{\mathrm{qp};i}(V_{{\rm s}0})\right]\cos(\omega_{0}t) 
\\ 
&&{}+ \frac{eV_{{\rm r}0}}{\hbar \omega_{0}}\left[\dot{Q}_{\mathrm{qpr};i}(V_{{\rm s}0} +  \hbar \omega_{0}/e )-\dot{Q}_{\mathrm{qpr};i}(V_{{\rm s}0})\right]\sin(-\omega_{0}t),
\nonumber 
\end{eqnarray}
where $\varphi_{\rm s}(t) = 2e/\hbar\int^{t} V_{\rm s}(t') {\rm d}t'$ is a slow-varying phase. \\

\section{Numerical results}
\label{sec:numerics}

Hereafter, $T_{\rm hot}$ denotes either $T_{\rm L}$ or $T_{\rm R}$ depending on the flux direction; we consider dependencies on the temperature ratio $T_{\rm hot}/T_{\rm c,L}$, with $T_{\rm c,L}$ being the critical temperature of the left superconductor of the SIS junction. Unless otherwise specified, we set the initial superconducting phase $\varphi_0=0$. Focusing first on the slowly-varying voltage applied to the SIS junction, we assume the form: $V_{\rm s}(t)=V_{{\rm s}0}\sin\Omega t$, with $V_{{\rm s}0} = 1~\mu$V and $\Omega = 1.5\pi$ GHz as numerical parameters for illustrative purposes. A similar frequency range was used in an experiment with a superconducting artificial atom, which includes a Josephson junction as a component~\cite{senior2020heat}. With $\Delta$ typically in the $10^{-4}$ to $10^{-3}$ eV range \cite{Kittel1976}, the condition $\hbar\Omega \ll \Delta$ is satisfied with $\Omega$ in the GHz range.

\begin{figure}[h]
  \centering
  \includegraphics[width=.5\textwidth]{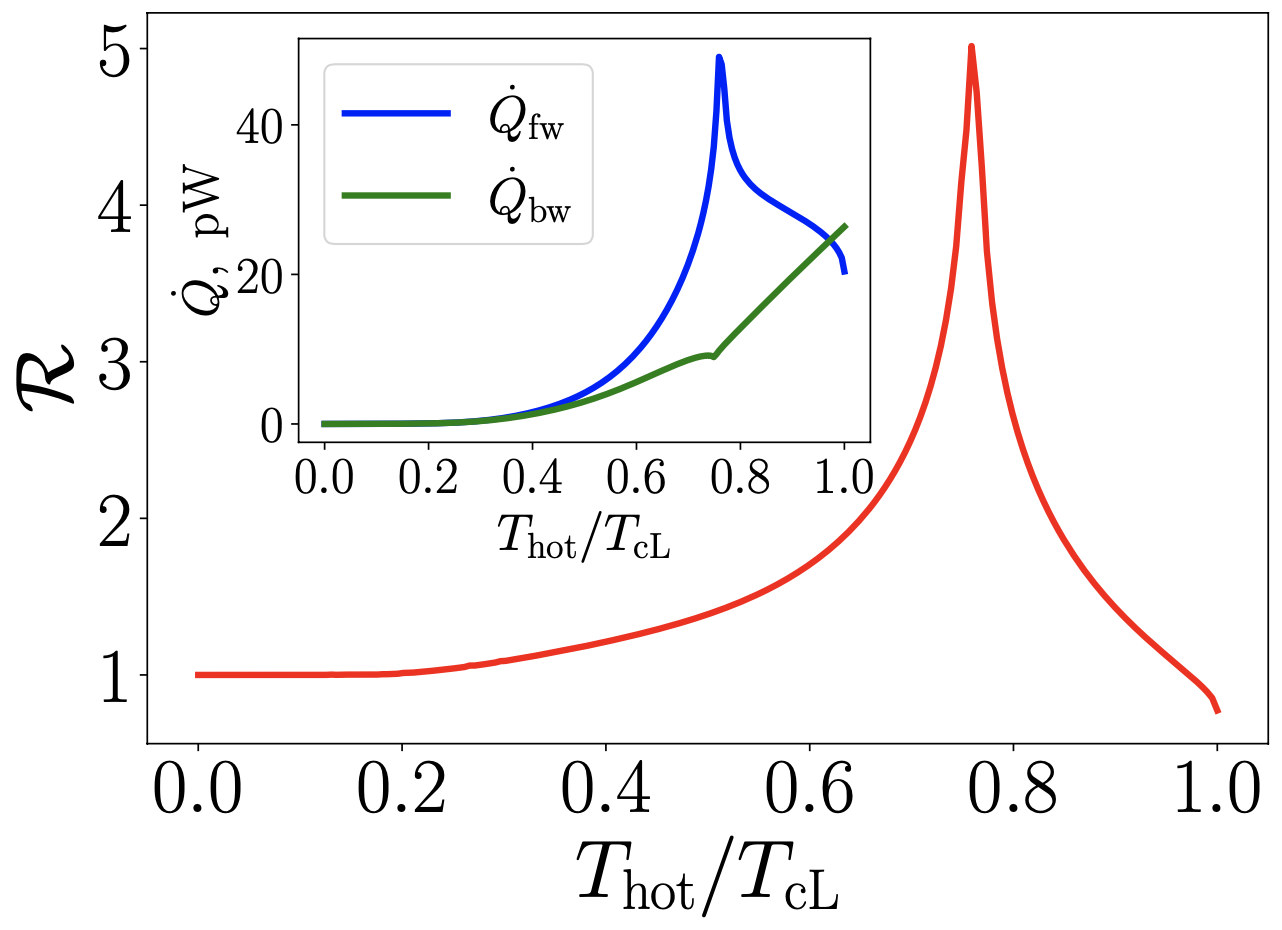}
  \caption[Heat fluxes]{Rectification coefficient and heat fluxes (insert) averaged over time as functions of the temperature ratio $T_{\rm hot}/T_{\rm cL}$ in the adiabatic regime. All the curves are plotted for $T_{\rm R} = 0.01T_{\rm cL}$, $\Omega = 1.5 \pi$ GHz, $V_{{\rm s}0}=1 \mu$V, and gap ratio $\Delta_{\rm R}(T_{\rm R}=0)/\Delta_{\rm L}(T_{\rm L}=0)=0.75$.} 
  \label{figfour}
\end{figure}

Figure~\ref{figfour} shows the heat rectification coefficient, defined as the ratio of the time-averaged forward and backward current, as well as the forward and backward heat fluxes against the temperature ratios $T_{\rm hot}/T_{\rm c,L}$. Note that since we consider an alternating voltage the Peltier heat contribution to the total heat flux is zero after time-averaging. We see that applying a slowly varying alternating voltage yields no large rise of ${\mathcal R}$ by comparison with the results of \cite{martinez2014coherent, fornieri2015electronic}, where the authors considered stationary heat currents with the same values of parameters as we did, but without voltage. Indeed, the forward and backward heat fluxes under the applied bias $V_{\rm s}$ have approximately the same values as those without voltage. This may be explained by the influence of the small variation of the superconducting phase $\varphi$ on the contributions to the heat fluxes; see Eq.~\eqref{eq:1}. As illustrated in Appendix~\ref{AppB}, the smallest contribution to the total heat flux is related to the tunneling of Cooper pairs (with $\sin\varphi\sim 0$), while the largest is associated with the quasiparticles' tunneling; the interference contribution (with $\cos\varphi\sim 1$) is comparable to the quasiparticles counterpart. 

\begin{figure}[h]
  \centering
  \hspace*{-.5cm}
  \includegraphics[width=.5\textwidth]{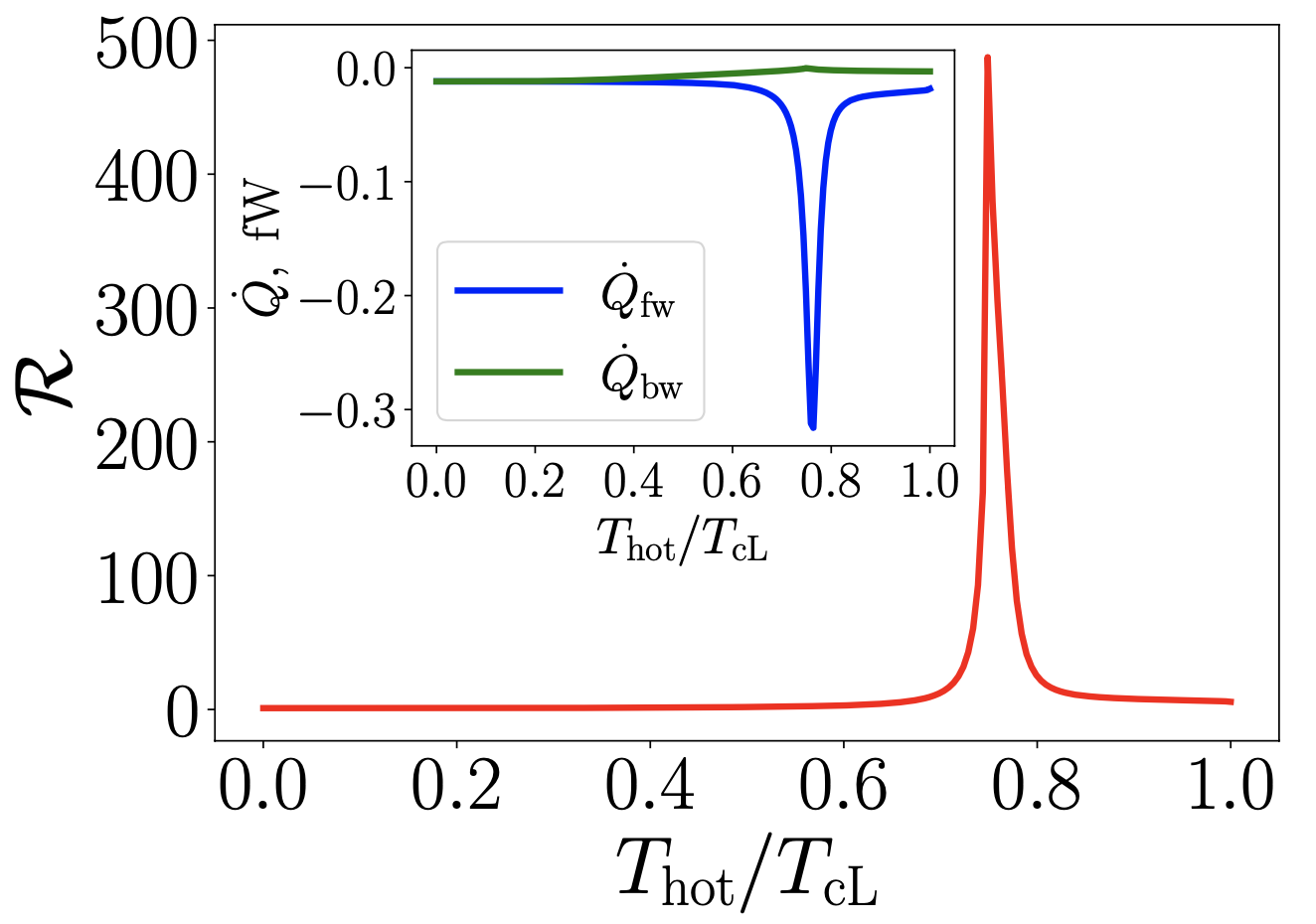}
  \caption{Rectification coefficient and heat fluxes (inset) averaged over time as functions of the temperature ratio $T_{\rm hot}/T_{\rm cL}$ for the non-adiabatic regime. All the curves are plotted for $T_{\rm R} = 0.01T_{\rm cL}$, $V_{{\rm s}0} = 1 \mu$V, $\Omega = 1.5 \pi$ GHz, $\omega_0 = 1.5 \pi$ THz, $V_{{\rm r}0}=10 \mu$V, and gap ratio $\Delta_{\rm R}(T_{\rm R}=0)/\Delta_{\rm L}(T_{\rm L}=0)=0.75$.}
  \label{figset}
\end{figure}  

For the rapidly-varying voltage, we choose a frequency $\omega_0$ that violates the condition of the adiabatic regime, with $\omega_0 \sim \Delta / \hbar \sim 1$ THz. To get such frequencies in experiments, one can use lasers \cite{ramian1992new}, which can be part of more complicated setups \cite{neu2018tutorial, lu2019controlled}. Figure~\ref{figset} shows the heat rectification coefficient and the forward as well as the backward heat fluxes beyond the adiabatic regime, after averaging over the period associated with the rapidly varying frequency, with the external applied bias given by the sum of the slow and rapid varying potentials: $V(t)=V_{\rm r}(t)+V_{\rm s}(t)$. Comparing our results with those of \cite{martinez2014coherent, fornieri2015electronic} where $\mathcal{R} \approx 10$, we see that going beyond the adiabatic regime in the THz range yields a sharp, 50-fold increase with $\mathcal{R} \approx 500$. 
The large peak in Fig.~\ref{figset} is due to the matching singularities in the superconducting density of states. The substantial increase of the rectification coefficient $\mathcal{R}$ compared to the adiabatic case is due to the additional contribution $\dot{Q}_{\mathrm{qpr}}$, which yields a drastic rise of the rectification coefficient because it amplifies the forward heat flux (see Appendix~\ref{AppB}). Note that in our simulation, the amplitude of the heat fluxes across the SIS junction is much smaller than in the adiabatic regime as the condition $eV_{{\rm r}0}/(\hbar \omega_0) \ll 1$, must be satisfied to ensure the validity of Eq.~\eqref{eq:8}~\cite{harris1975josephson}. Here, for numerical calculations, $eV_{{\rm r}0}/(\hbar \omega_0) \approx 10^{-2}$. We checked that the large increase of $\mathcal{R}$ remains the same under a ten-fold increase of the voltage amplitude.

\begin{figure}[h]
  \centering
  \includegraphics[width=.5\textwidth]{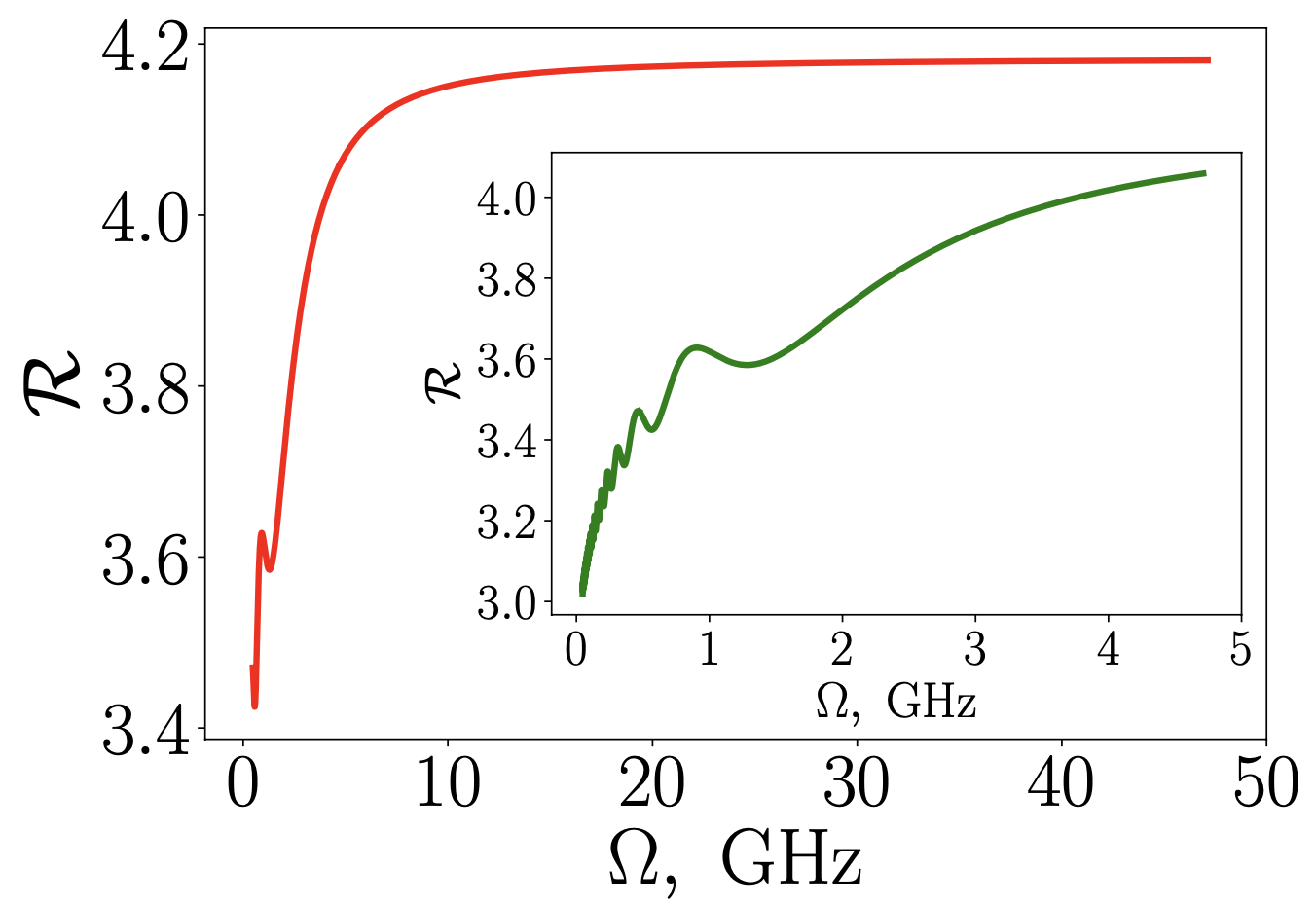}
  \caption{Rectification coefficient averaged over time as a function of the frequency $\Omega$ in the adiabatic regime and a magnification of the main plot at small frequencies (inset). We plotted all the curves for $T_{\rm R} = 0.01T_{\rm cL}$, gaps' ratio  $\Delta_{\rm R}(T_{\rm R}=0)/\Delta_{\rm L}(T_{\rm L}=0)=0.75$, the temperature ratio $T_{\rm hot}/T_{\rm cL}=0.77$, and $V_{{\rm s}0} =1 \mu$V.}
  \label{figde}
\end{figure}

\begin{figure}[h]
  \centering
  \includegraphics[width=.5\textwidth]{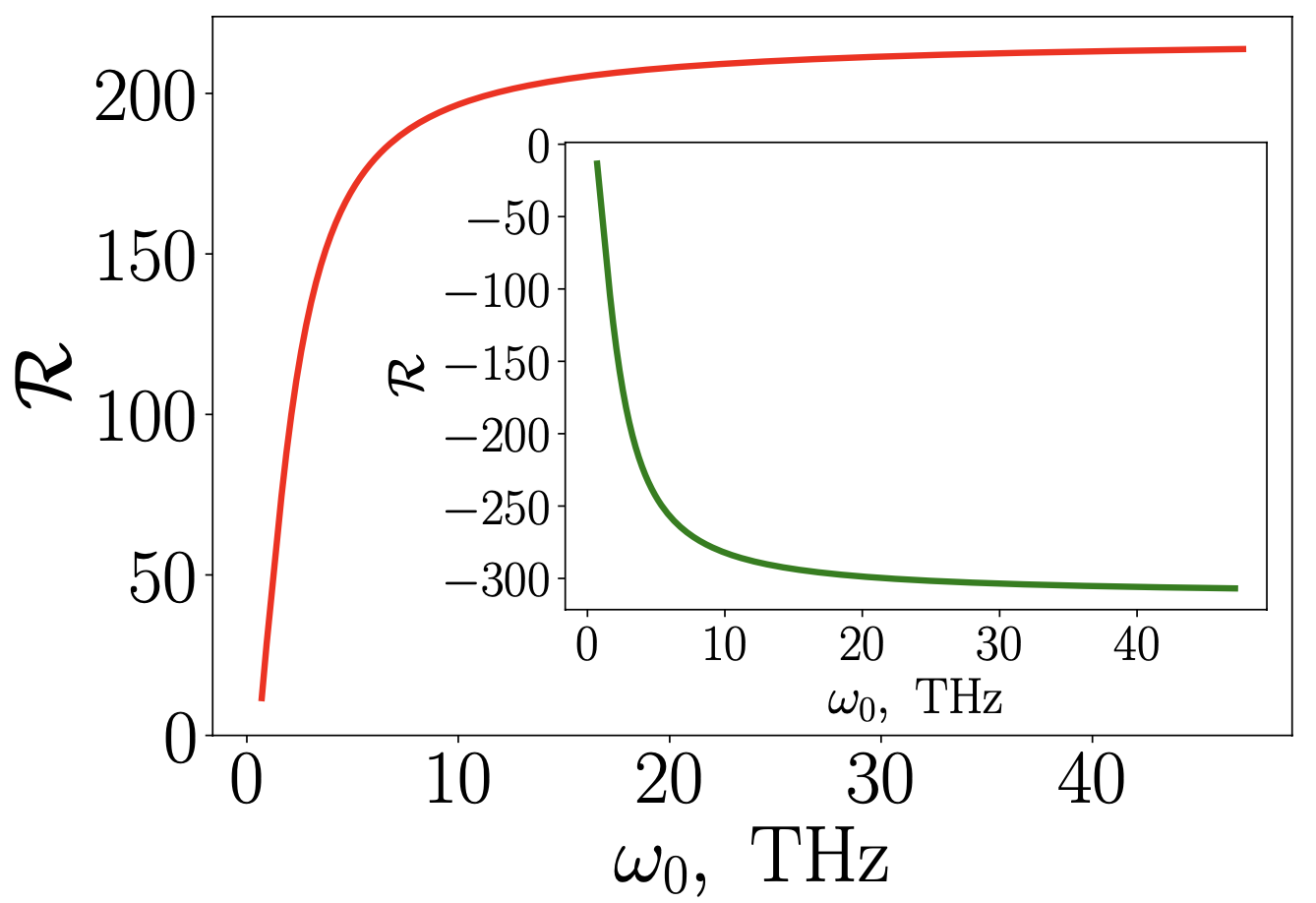}
  \caption{Rectification coefficient averaged over time as a function of the frequency $\omega_{0}$, depicted for the non-adiabatic regime. The inset graph displays the rectification coefficient for the initial phase $\varphi_0 = \pi$.  We plotted all the curves for $T_{\rm R} = 0.01T_{\rm cL}$, gaps' ratio  $\Delta_{\rm R}(T_{\rm R}=0)/\Delta_{\rm L}(T_{\rm L}=0)=0.75$, the temperature ratio $T_{\rm hot}/T_{\rm cL}=0.77$, $V_{{\rm s}0} =1 \mu$V, $\Omega = 1.5 \pi$ GHz  and $V_{{\rm r}0} =1 \mu$V.}
  \label{figtr}
\end{figure}

The heat rectification coefficient does not increase drastically by comparison with \cite{martinez2014coherent, fornieri2015electronic}, in the adiabatic regime, i.e. for frequencies up to the GHz range, while it does beyond, in the THz range. So it is instructive to analyze the behavior of the rectification coefficient in a SIS junction as a function of frequency. Figures ~\ref{figde} and \ref{figtr} depict behaviors that are quite similar qualitatively, but that differ by almost two orders of magnitude.

In the megahertz range (insert in Fig.~\ref{figde}), we observe oscillations that show the low-frequency response of the Josephson junctions to an alternating voltage resulting in alternating heat fluxes.
These oscillations are due to the tunneling and interference contributions to the heat flow, which depend on the phase. In turn, the phase depends on time, see Eq.~\eqref{eq:1}, as $\varphi(t)=\varphi_0-\frac{2eV_{{\rm s}0}}{\hbar\Omega}\cos(\Omega t)$ (modulo $2\pi$), so that averaging over time spans all possible values of the phase when $\frac{2eV_{{\rm s}0}}{\hbar\Omega}=2 n\pi$, with $n$ integer. The phase periodicity is thus reflected in oscillations of ${\mathcal R}$, with peaks at frequencies proportional to $1/n$.

In contrast, as $\Omega > 1$ GHz, the rectification coefficient experiences a rather sharp monotonic increase up to $\sim$ 10 GHz before saturation in the adiabatic regime. Beyond the adiabatic regime, no oscillation is observed: ${\mathcal R}$ rises monotonically across two orders of magnitudes over several THz. Note that in the insert of Fig.~\ref{figtr}, numerical results are shown for an initial phase $\varphi_{0}= \pi$, while the main plot is drawn for $\varphi_{0}= 0$. Since the sign of the rectification 
coefficient is positive for $\varphi_{0}= 0$ and negative for $\varphi_{0}= \pi$, we can conclude that the initial phase $\varphi_{0}$ can act like a control knob.

\section{Discussion}
\label{sec:discussion}

A thermal diode based on an SIS junction, can be implemented like in \cite{martinez2014coherent, martinez2015rectification} with an externally applied oscillating voltage.  
Heat rectification under the presence of constant voltage was investigated in ferromagnetic insulator-based superconducting tunnel junctions \cite{giazotto2020very}, where a huge rectification coefficient was reported  $\approx 500$ (owing to the combined effects of spin splitting and spin polarization) even though there was only one electronic (quasiparticle) contribution to the heat current because of the different nature of metals and ferromagnetic materials. The superconducting phase played no role in that case, i.e., the coherent effects were ignored. In our case, instead, the phase can be used to tune amplitude and sign of the rectification effect. The voltage rise in that work was related to the temperature gradient because of thermoelectric coupling \cite{Apertet2016}, while here we consider both a temperature gradient and an external voltage bias. Beyond the adiabatic regime, a SIS junction might serve as a heat valve if the applied voltage has a slowly-varying part and a rapidly-varying part as the rectification coefficient is much different from 1 only when $\Delta_{\rm L}(T_{\rm L}) = \Delta_{\rm R}(T_{\rm R})$. This could open a new avenue in the use of heat valves due to the involvement of Cooper pairs in addition to existing realizations \cite{ronzani2018tunable, dutta2020single}.\\

\section{Concluding remark}
\label{sec:conc}

We theoretically studied heat rectification in a SIS junction considering an external applied voltage in addition to a temperature bias. To quantitatively characterize the combined effect of voltage and temperature on the heat rectification, we calculated all the contributions to the forward and backward heat fluxed under non-zero voltage in the adiabatic (or low frequency -- GHz) regime, and beyond (high-frequency -- THz). We found that a rapidly-varying potential causes a large enhancement of the heat rectification coefficient due to the interference term in the heat current $\dot Q_{\mathrm{int}}$. An analysis of heat fluxes has been conducted to find out the nature of such a drastic rise in heat rectification. Our theoretical work can be implemented in experiments. Fabrication of efficient, yet complex hybrid systems \cite{martinez2015rectification,giazotto2020very} can be replaced by or complemented with much simpler superconducting circuits components under an externally applied voltage. \\

\emph{Acknowledgments --}
We are pleased to thank Konstantin Tikhonov for fruitful discussions. I.K. and G.B. acknowledge support by the INFN through the project QUANTUM.\\

\begin{figure}[H]
  \includegraphics[width=.4\textwidth]{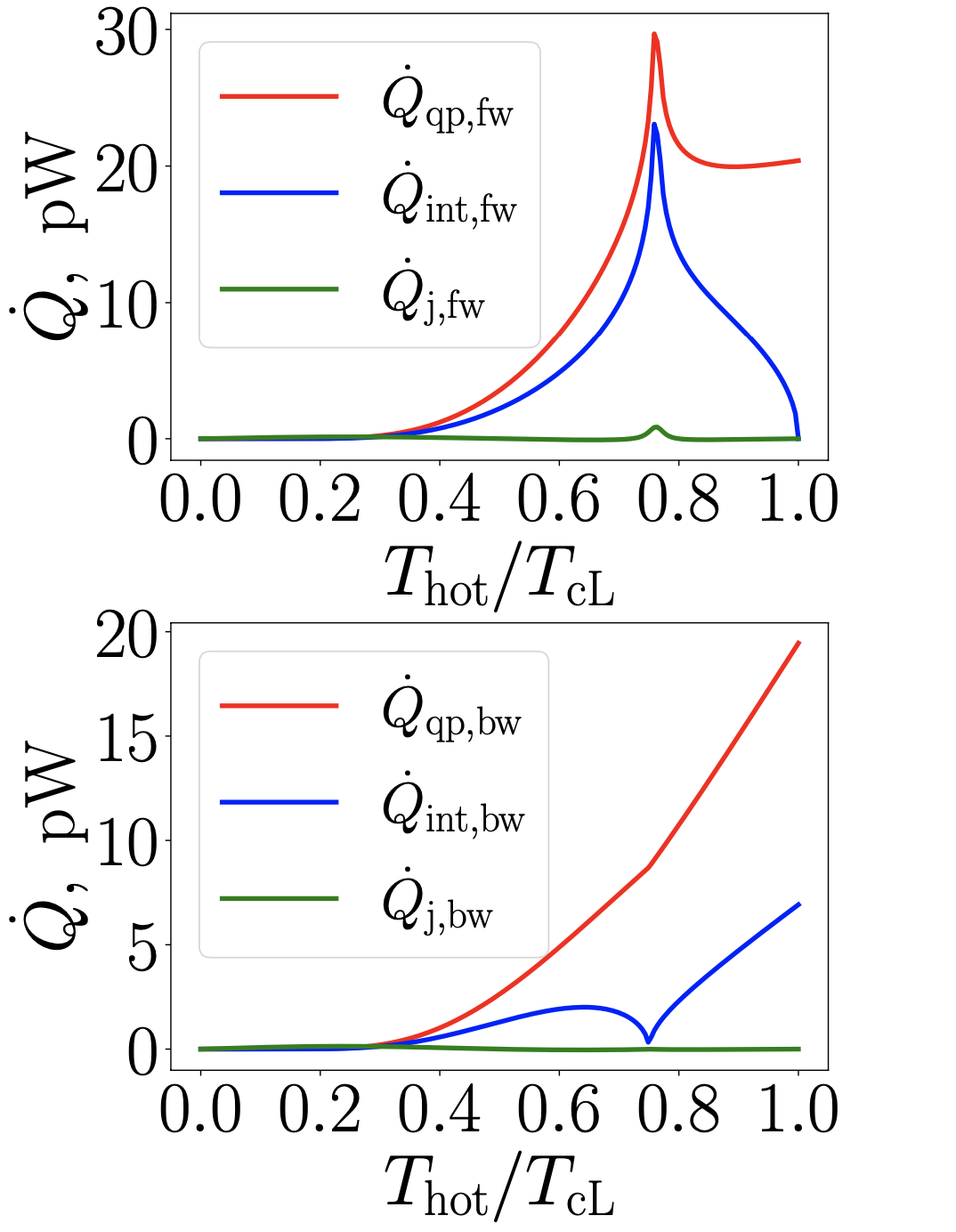}
  \caption[Heat fluxes]{Contributions to the total heat flux as  function of the temperature ratio $T_{\rm hot}/T_{\rm cL}$. The plots show the time-averaged heat fluxes $\langle\dot{Q}_{\mathrm{qp}}\rangle_{t}$, $\langle\dot{Q}_{\mathrm{int}} \mathrm{cos}(\varphi)\rangle_{t}$, $\langle\dot{Q}_{\rm j} \mathrm{sin}(\varphi)\rangle_{t}$: upper panel -- forward; bottom panel -- backward. We plotted all the curves for $T_{\rm R} = 0.01T_{\rm cL}$ and gaps' ratio $\Delta_{\rm R}(T_{\rm R}=0)/\Delta_{\rm L}(T_{\rm L}=0)=0.75$,
  $V_{{\rm s}0} =1 \mu$V, $\Omega = 1.5\pi$ GHz.}
  \label{figfive}
\end{figure}

\begin{figure}[H]
  \includegraphics[width=.4\textwidth]{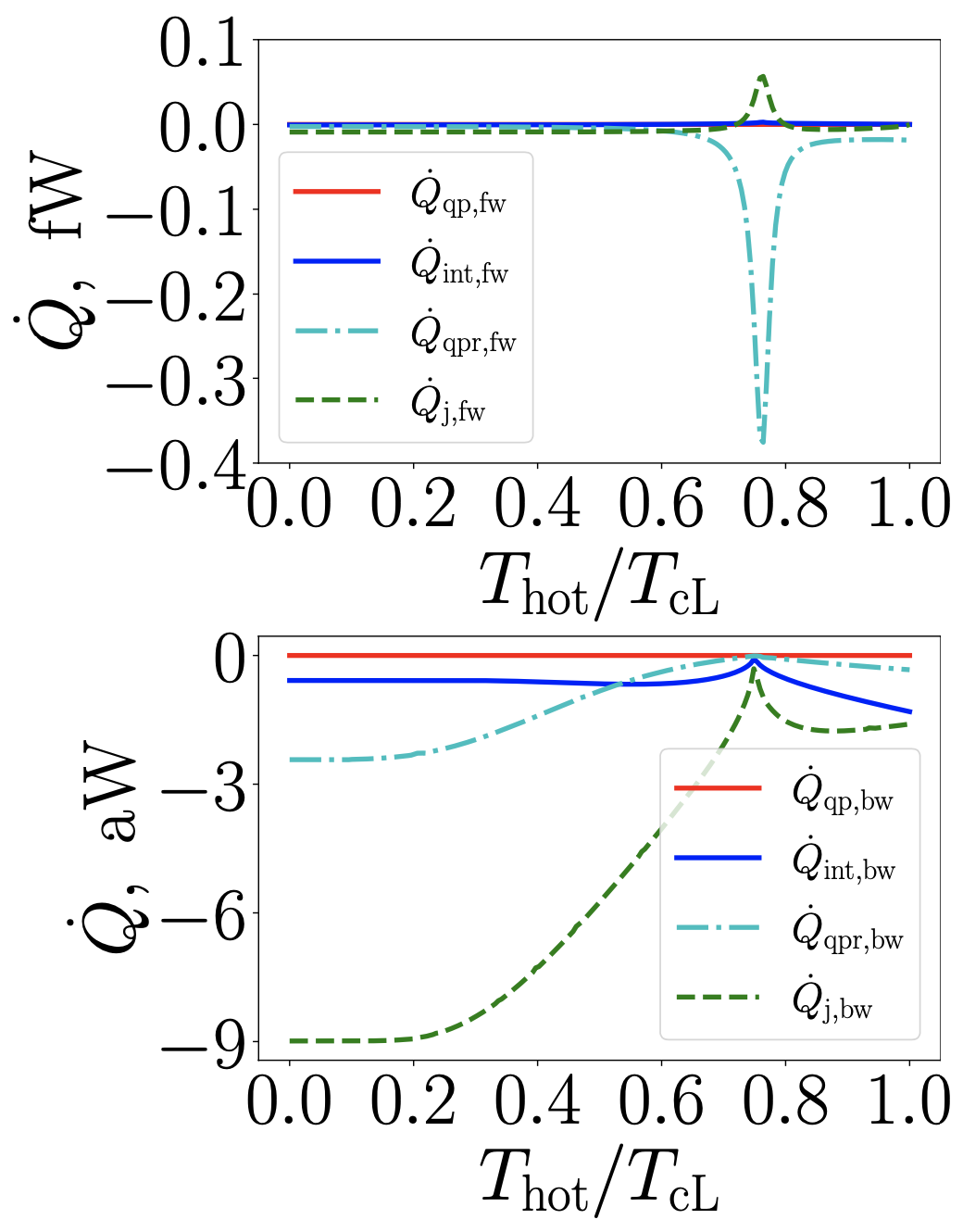}
  \caption{Time-averaged contributions to the total heat flux as a function of the temperature ratio $T_{\mathrm{hot}}/T_{\rm cL}$. The upper plot corresponds to the forward heat fluxes, while the bottom one corresponds to the backward heat fluxes. We plotted all the curves for $T_{\rm R} = 0.01T_{\rm cL}$, gaps' ratio  $\Delta_{R}(T_{\rm R}=0)/\Delta_{\rm L}(T_{\rm L}=0)=0.75$, $V_{{\rm s}0} = 10 \mu$V, $\Omega = 1.5 \pi$ GHz, $V_{{\rm r}0} = 10 \mu$V, and $\omega_{0} = 1.5 \pi$ THz.}
  \label{fignv}
\end{figure}

\begin{widetext}
\appendix
\section{Microscopic expressions of the heat currents}
\label{AppA}

We provide the general expressions for each component of the heat fluxes in Eqs.~(\ref{eq:3}) \cite{marchegiani2020nonlinear}.

\begin{equation}
Q_{\mathrm{qp};i} = \frac{G}{e^{2}} \int_{- \infty}^{\infty} E_{k} \ \mathrm{Im}[N_{k}(E_{k})]\mathrm{Im} [N_{m}(E_{m})][f_{k}(E_{k})-f_{m}(E_{m})]~{\rm d}E,
\label{eq:A1}
\end{equation}
\begin{equation}
Q_{{\rm j};i} = \frac{G}{2e^{2}} \int_{- \infty}^{\infty} E_{k} \bigg \{ \mathrm{Im}[F_{k}(-jE_{k})]\mathrm{Re}[F_{m}(jE_{m})]\mathrm{tanh} \left( \frac{E_{k}}{2k_{\rm B}T_{k}} \right) + \mathrm{Im}[F_{m}(jE_{m})]\mathrm{Re}[F_{k}(-jE_{k})] \mathrm{tanh} \bigg ( \frac{E_{m}} {2k_{\rm B}T_{m}} \bigg) \bigg \}~{\rm d}E,
\label{eq:A2}
\end{equation}
\begin{equation}
Q_{\mathrm{int};i} = \frac{G}{e^{2}} \int_{- \infty}^{\infty} E_{k} \ \mathrm{Im}[F_{k}(-jE_{k})]\mathrm{Im}[F_{m}(jE_{m})][f_{k}(E_{k})-f_{m}(E_{m})]~{\rm d}E,
\label{eq:A3}
\end{equation}
\begin{equation}
Q_{\mathrm{qpr};i} = \frac{G}{2e^{2}} \int_{- \infty}^{\infty} E_{k} \bigg \{ \mathrm{Im}[N_{k}(E_{k})]\mathrm{Re}[N_{m}(E_{m})]\mathrm{tanh} \left( \frac{E_{k}}{2k_{\rm B}T_{k}} \right) + \mathrm{Im}[N_{m}(E_{m})]\mathrm{Re}[N_{k}(E_{k})] \mathrm{tanh} \bigg ( \frac{E_{m}} {2k_{\rm B}T_{m}} \bigg) \bigg \}~{\rm d}E,
\label{eq:A4}
\end{equation}
where $\mathrm{Re[\ldots]}$ and $\mathrm{Im[\ldots]}$ denote the real and imaginary parts, $j^2 = -1$, $i \equiv {\rm fw}$ or ${\rm bw}$, $m = {\rm R} $ and $k = {\rm L}$ for the forward flux and vice versa for the backward flux, and $ E_k= E - \mu_k$ with $\mu_k$ being the electrochemical potential \cite{barone1982physics, marchegiani2020phase}. According to the BCS model, the quasiparticle densities of states are given by $N_{k}(E_{k})= -(E_{k}+j\Gamma_{k})/\sqrt{\Delta_{k}(T)^{2}-(E_{k}+j\Gamma_{k})^{2}}$, and the anomalous Green functions by $F_{k}(-jE_{k})=\Delta_{k}(T)/\sqrt{\Delta_{k}(T)^{2}-(E_{k}+j\Gamma_{k})^{2}}$, where $\Gamma_{k} = 10^{-4}\Delta_{k}(0)$ are the Dynes parameters, $\Delta_{k}(T_k)$ are the superconducting gaps, and $G=10^{-3}~\Omega^{-1}$ is the normal state electrical conductance of the junction.

\section{Contributions to heat fluxes}
\label{AppB}
In this appendix, we show how several flux components behave in different regimes and how they influence heat rectification. Let us start with the adiabatic regime, for which all the contributions are depicted in Fig.~\ref{figfive}. The dominating contribution for temperature ratios $T_{\rm hot}/T_{\rm cL} \geq 0.4$ is the quasiparticle heat currents. The interference contribution is next, while the smallest is the Cooper pair contribution. This is why there is no steep increase in the rectification coefficient in the adiabatic regime compared to a situation, when a voltage is applied~\cite{fornieri2015electronic}.

To gain insights on the rise of the rectification coefficient beyond the adiabatic regime, we plot each of the contributions to the total heat flux when the applied voltage has a slowly and a rapidly varying part, in Fig.~\ref{fignv}. For the forward heat fluxes, the reactive contribution of the quasiparticle current is the most significant, albeit near its peak the Cooper pairs' part of the quasiparticle heat current is not negligible; so these contributions mostly determine how each of the heat fluxes behave. While the forward components have different signs, the backward components are non positive. This suggests that such a junction may be used as a heat valve. Indeed, to block the forward heat flux, we can reduce the temperature of the hot electrode close to zero, whereas if it is needed to amplify the heat current, one can increase this temperature to the point where the peak of the heat fluxes occurred.
\end{widetext}

\newpage

~\\

\vskip 1cm 

\bibliography{main}
\end{document}